\pdfoutput=1
\documentclass[%
reprint,
superscriptaddress,
amsmath,
amssymb,
aps,floatfix]{revtex4-2} 
\usepackage{times}
\usepackage{amsmath}
\usepackage{float}
\usepackage{bm}
\usepackage{gensymb}
\usepackage[dvipsnames]{xcolor}
\usepackage{graphicx}
\usepackage{amssymb}
\usepackage{url,hyperref}

\DeclareUnicodeCharacter{2009}{\,}
\newcommand{\Dmin}{D_\text{min}^2}

\usepackage{xcolor,hyperref}
\hypersetup{
   colorlinks,
   linkcolor={blue!50!black},%{red!80!black},
   citecolor={blue!50!black},
   urlcolor={blue!80!black}
}
\newcommand{\mlm}[1]{\textcolor{black}{#1}}
\newcommand{\tah}[1]{\textcolor{black}{#1}}

\begin{document}

\title{Using the force landscape of an active solid to predict plastic deformation}

\author{Tyler Hain}
%\email[]{tahain@syr.edu}
\affiliation{Department of Physics and BioInspired Institute, Syracuse University, Syracuse NY 13210}

\author{Edan Lerner}
\email[]{e.lerner@uva.nl}
\affiliation{Institute for Theoretical Physics, University of Amsterdam, Science Park 904, 1098 XH Amsterdam, the Netherlands}

\author{M. Lisa Manning}
\email[]{mmanning@syr.edu}
\affiliation{Department of Physics and BioInspired Institute, Syracuse University, Syracuse NY 13210}

\begin{abstract}
Non-active disordered solids feature quasilocalized excitations that control plasticity, similar to crystal lattice defects, and these excitations can be identified via harmonic or anharmonic analyses of the potential energy landscape. Here we explore whether such ideas can be extended to active matter, focusing on dense packings of self-propelled rods. We generalize the definition of nonlinear excitations to force landscapes that incorporate active, non-conservative forces and find that force-based cubic excitations robustly predict future plastic events, enabling control of active solids.
\end{abstract}

\maketitle

\noindent
\textbf{Introduction.} Engineers want to design functional materials that sense stimuli and reconfigure to perform tasks, while life scientists are interested in explaining how such processes are controlled in living organisms.  These functions require targeted mechanical responses such as controlled flows, shape changes, and altered rheological properties. Active matter systems, where the constitutive particles consume energy, are a useful paradigm for such systems as they are out of equilibrium~\cite{ramaswamy_mechanics_2010} and controllable~\cite{reyes_garza_magnetic_2023}.

Active matter cannot be described in terms of an effective potential or Hamiltonian, abrogating key techniques of condensed matter and statistical physics; nevertheless, much progress has been made. Initial research focused on low- or intermediate-density regimes exhibiting motility-induced phase separation \cite{fily_athermal_2012, cates_motility-induced_2015} and odd viscosity \cite{banerjee_odd_2017, souslov_topological_2019, fruchart_odd_2023} or on active solids with a crystalline structure \cite{bililign_motile_2022, vansaders_sculpting_2021, tan_odd_2022, scheibner_odd_2020, baconnier_selective_2022}.

Recent work has focused on dense and disordered active systems, which inherit many features of glassy materials, including intermittent plasticity, history and protocol dependence, and avalanches \cite{berthier_yielding_2025, mandal_extreme_2020, mandal_shear-induced_2021, morse_direct_2021}. Activity strongly impacts the glass transition \cite{berthier_non-equilibrium_2013, berthier_how_2017, nandi_random_2018, janssen_active_2019, debets_cage_2021, paul_dynamical_2023, priya_active_2025, shee_tuning_2025}, critical scaling around the jamming point, \cite{henkes_active_2011, liao_criticality_2018, anand_active_2024}, and emergent long-range velocity correlations \cite{szamel_long-ranged_2021, henkes_dense_2020}.  Dense active solids exhibit new dynamical regimes with emergent ordering \cite{mandal_shear-induced_2021} and distinct yielding behavior \cite{villarroel_critical_2021,villarroel_avalanche_2024, mandal_extreme_2020,ghaznavi_yielding_2025}, which is reasonably well-captured by mesoscale elastoplastic models when the activity is small~\cite{ghosh_elastoplastic_2025} or highly persistent~\cite{ghaznavi_yielding_2025}. However, elastoplastic models are not yet quantitatively predictive, as one cannot extract key model parameters from active simulations or experiments.

In non-active amorphous materials, plasticity is known to occur in localized bursts at ``soft spots" that are the analogue of lattice defects in crystalline systems and correspond to quasilocalized vibrational modes of the potential energy landscape \cite{maloney_amorphous_2006, lerner_statistics_2016, xu_anharmonic_2010, falk_dynamics_1998, chen_measurement_2011, manning_vibrational_2011, stanifer_avalanche_2022, richard_mechanical_2023, berthier_yielding_2025}. Methods for identifying these soft spots include a linear~\cite{manning_vibrational_2011} or nonlinear~\cite{kapteijns_nonlinear_2020} analysis of the landscape, machine learning methods trained on the structure~\cite{cubuk_identifying_2015, bapst_unveiling_2020}, direct application of strain~\cite{ruan_predicting_2022}, and others~\cite{richard_predicting_2020}. Once the statistics of these defects are known, they immediately predict mechanical and thermal properties~\cite{degiuli_effects_2014, moriel_wave_2019, wang_sound_2019, ji_thermal_2020, wang_low-frequency_2019, rainone_pinching_2020, richard_brittle--ductile_2021} and specify key model parameters in highly successful elastoplastic constitutive laws~\cite{nicolas_deformation_2018, zhang_structuro-elasto-plasticity_2022, xiao_identifying_2023}.

In active systems, mechanical instabilities leading to rearrangements are governed by both the potential energy landscape of standard reciprocal interactions \emph{and} the non-potential field of active or non-reciprocal forces. Is a defect-based statistical description even the correct starting point? One can design special active matter systems that map onto an effective potential landscape; linear landscape analysis methods fail to predict plasticity due to a strong coupling between vibrational mode frequencies and heterogeneous pressure~\cite{giannini_searching_2022}, while nonlinear methods account for this coupling and perform better~\cite{giannini_defects_2024}. In active crystals, defects can be used to tune the material's shape \cite{vansaders_sculpting_2021, bililign_motile_2022}. Are defects still predictive of plasticity in generic disordered active solids? If so, how do we identify them?

Here we generalize potential-landscape methods to generate force-based definitions of quasilocalized excitations \cite{kapteijns_nonlinear_2020}, use these to locate defects in a numerical model of active self-propelled rods \cite{wensink_meso-scale_2012, grauer_spontaneous_2018, janssen_aging_2017}, and study their correlations with plastic events.

\noindent
\textbf{Methods.}  To identify defects, we extend potential energy landscape methods from passive solids where vibrations around a metastable minimum are hybridizations between collective phononic wave-like modes with a Debye frequency spectrum $\omega^{d-1}$, and quasilocalized modes with a frequency distribution growing as $\omega^4$ \cite{lerner_low-energy_2021}. Quasilocalized modes are associated with defects in the local structure~\cite{wijtmans_disentangling_2017, richard_predicting_2020}.

One method to disentangle these hybridized modes and find defects relies on nonlinear properties of the potential energy landscape~\cite{gartner_nonlinear_2016, gartner_nonlinear_2016-1,  lerner_micromechanics_2016, kapteijns_nonlinear_2020}. To define nonlinear modes, we begin with the energy function $U(\{x_i\})$ and degrees of freedom $\{x_i\}$. The energy cost of a perturbation $\delta \mathbf{x} = s\mathbf{\hat{z}}$ away from a local minimum to third order in $s$ is $\delta U\! =\! (1/2)\kappa(\mathbf{\hat{z}})s^2\! +\! (1/6)\tau(\mathbf{\hat{z}}) s^3\! +\! \mathcal{O}(s^4)$, where $\mathbf{\hat{z}}$ is a unit vector in phase space denoting the direction of the perturbation and $s$ is the magnitude of the displacement. The coefficients $\kappa(\mathbf{\hat{z}})$ and $\tau(\mathbf{\hat{z}})$ describe the curvature and asymmetry, respectively, of the energy landscape along the direction $\mathbf{\hat{z}}$. 

A cubic approximation of the energy barrier along the direction $\mathbf{\hat{z}}$ is  $b(\mathbf{\hat{z}}) \equiv (2/3) \kappa(\mathbf{\hat{z}})^3/\tau(\mathbf{\hat{z}})^2$. Previous work has demonstrated that the displacement modes that minimize $b(\mathbf{\hat{z}})$, termed nonlinear plastic modes, are quasilocalized and correlated with future plastic rearrangements \cite{gartner_nonlinear_2016, gartner_nonlinear_2016-1, lerner_micromechanics_2016, kapteijns_nonlinear_2020}.

To generalize these nonlinear modes to systems with only a force law $\mathbf{f} = -\nabla U + \mathbf{f}^{\text{act}}$, where $U$ describes any conservative interactions and $\mathbf{f}^{\text{act}}$ are the non-conservative forces, we follow previous work~\cite{gartner_nonlinear_2016, kapteijns_nonlinear_2020, richard_mechanical_2023} and expand the force response to a perturbation as
\begin{align}
    \mathbf{f}(\mathbf{\hat{z}}, s) &= \mathbf{f}^{(1)}(\mathbf{\hat{z}})s + \frac{1}{2}\mathbf{f}^{(2)}(\mathbf{\hat{z}})s^2 + \mathcal{O}(s^3).
\end{align}
We use relationships between the coefficients to define special populations of excitations. In conservative systems, harmonic modes are displacements $\mathbf{\hat{\phi}}$ parallel to the resulting first-order force response $\mathbf{f}^{(1)}(\mathbf{\hat{\phi}}) \propto \mathbf{\hat{\phi}}$; we use this same equation to define a generalized set of 'harmonic modes' in active systems which are eigenvectors of the 'augmented Hessian' $\mathcal{M} = \nabla \mathbf{f}^{(1)}$. 

Similarly, standard nonlinear plastic modes have the property that the first-order and second-order coefficients of the force response are parallel, $\mathbf{f}^{(1)}(\hat{\pi}) \propto \mathbf{f}^{(2)}(\hat{\pi})$, and can be obtained by minimizing $b(\mathbf{\hat{z}})$ (see Supplemental Materials). Here we define generalized cubic modes in non-conservative systems as displacements $\mathbf{\hat{\pi}}$ that are solutions to the equation
\begin{align}
    \mathbf{f}^{(1)}(\mathbf{\hat{\pi}}) = \frac{\kappa(\hat{\pi})}{\tau(\hat{\pi}) }\mathbf{f}^{(2)}(\mathbf{\hat{\pi}}),
    \label{cubic mode equation}
\end{align}
where $\kappa(\mathbf{\hat{z}}) \equiv \mathbf{\hat{z}}\cdot\mathbf{f}^{(1)}(\mathbf{\hat{z}})$ and $\tau(\mathbf{\hat{z}}) \equiv \mathbf{\hat{z}}\cdot\mathbf{f}^{(2)}(\mathbf{\hat{z}})$. These quantities can still be interpreted as a stiffness and asymmetry, respectively, because they determine the work required to deform the system along the direction $\mathbf{\hat{z}}$ (See Supplemental Materials).  Therefore, $b(\mathbf{\hat{z}})\equiv(2/3)\kappa(\mathbf{\hat{z}})^3/\tau(\mathbf{\hat{z}})^2$ still represents something akin to an ``energy barrier", although this does not imply that there is a global energy landscape, as the work required to deform a non-conservative system is generally path-dependent.

\begin{figure}[h]
\centering
\includegraphics[scale= 0.1]{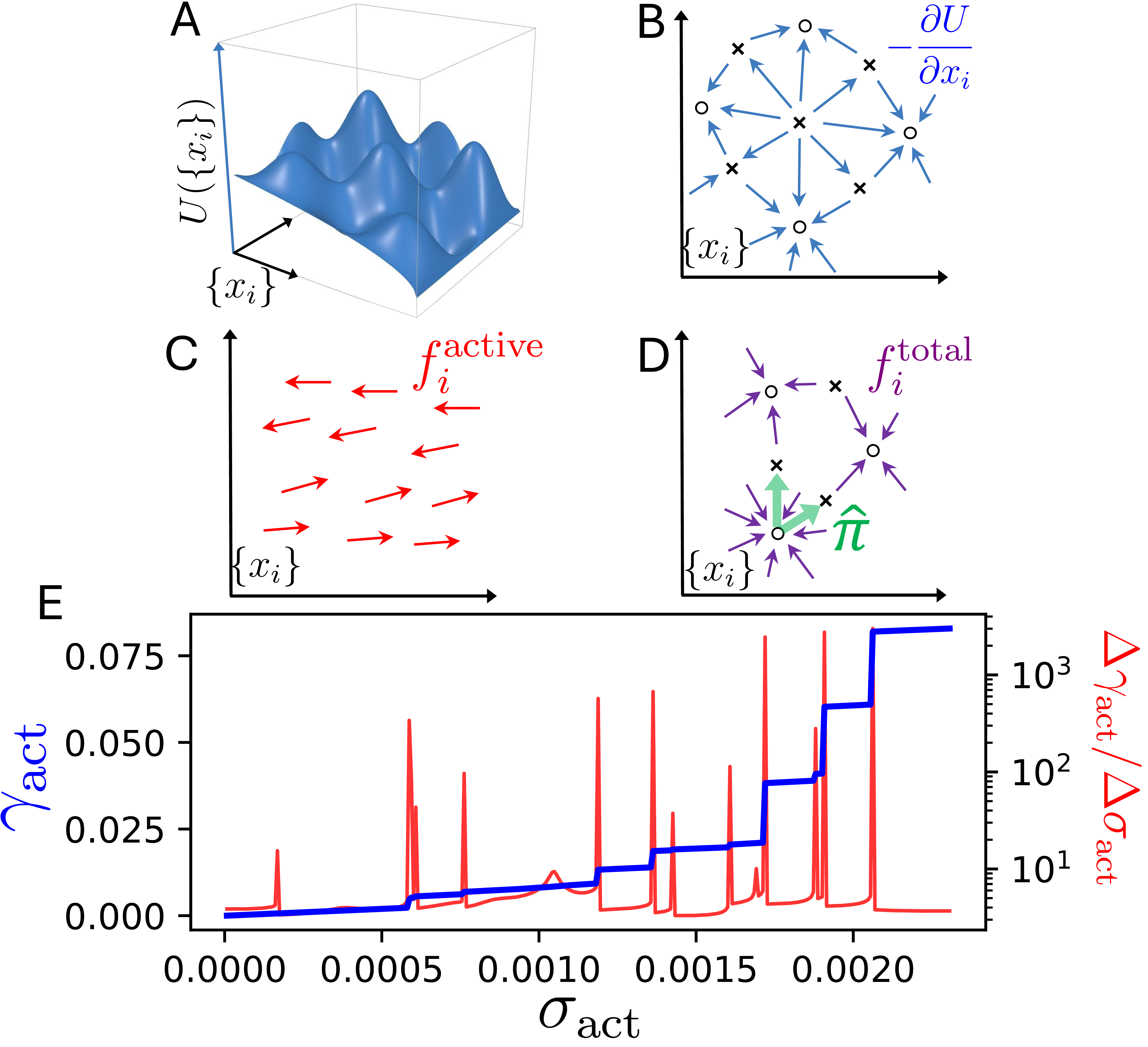}
\caption{A) Schematic of a potential energy landscape in a disordered packing, which can be represented as a force landscape B) with stable and unstable fixed points. C) A landscape of non-conservative, active forces. D) The total force landscape is the sum of (A) and (B) which will still have stable and unstable fixed points if active forces are not too large. Cubic modes (green) point from metastable states towards nearby instabilities. E) Plastic deformation, i.e. spikes in the strain per unit stress, $\Delta \gamma_{\text{act}}/ \Delta \sigma_{\text{act}}$ (red curve), occur when this landscape changes (via increasing non-conservative forces $\sigma_{\text{act}}$) until an unstable fixed point combines with the stable fixed point in a saddle-node bifurcation, causing the system relax into a nearby metastable state, i.e. a discontinuous jump in a strain-stress curve (blue).}
\label{ForceLandscapeCartoon_StressStrain}
\end{figure}

We apply these methods to simulations of 2D jammed packings of rod-shaped particles with self-propulsion forces directed along their long axes and periodic boundary conditions~\cite{wensink_meso-scale_2012, grauer_spontaneous_2018, janssen_aging_2017}. We focus on rods instead of spheres because in spheres the self-propulsion forces are extra degrees of freedom with their own dynamics~\cite{henkes_active_2011, mandal_extreme_2020, anand_active_2024}, and the packing has an 'instantaneous energy landscape' that evolves on the active force persistence timescale (or which doesn't evolve at all between plastic events for infinite persistence times \cite{ghaznavi_yielding_2025, gandikota_jammed_2026}). In rods the self-propulsion (orientated always along the long axis) is instead governed by contact-force induced torques, and is not an independent degree of freedom. Therefore, dense active rods cannot be described by an energy landscape over any well-defined timescale.

We model each rod as five constrained spherical sub-particles interacting via a pair-wise repulsive potential, as in previous work~\cite{wensink_meso-scale_2012, grauer_spontaneous_2018, janssen_aging_2017}. We generate ensembles of 120 packings of various sizes ($N=256, 512, 1024$) at a density $\rho = 5N/L^2 = 0.75$, which is far above the density at which the system jams, $\rho \simeq 0.6$. \mlm{To generate flow, each active force $\sigma_\text{act}$ starts at zero magnitude and is quasistatically and uniformly increased until the entire system begins to flow, a protocol that was also recently used to study plasticity near the jamming transition in systems of self-propelled spheres with infinite persistence \cite{gandikota_jammed_2026}. In contrast to 'athermal quasi-static random displacement' protocols that fix the strain between particles and measure the stress, akin to fixed strain rate rheology protocols~\cite{morse_direct_2021,ghaznavi_yielding_2025}, our protocol fixes the forces and measures the strain, akin to creep rheology protocols.} To compare our results directly to global shear~\cite{morse_direct_2021}, we write the stress due to the active forces as $\sigma_{\text{act}} = \eta/L\sqrt{12}$, where $\eta$ is the magnitude of the active force on each rod and $L$ is the linear system dimension (see Supplemental Materials). 

\begin{figure}[h!]
\centering
\includegraphics[scale= 0.1]{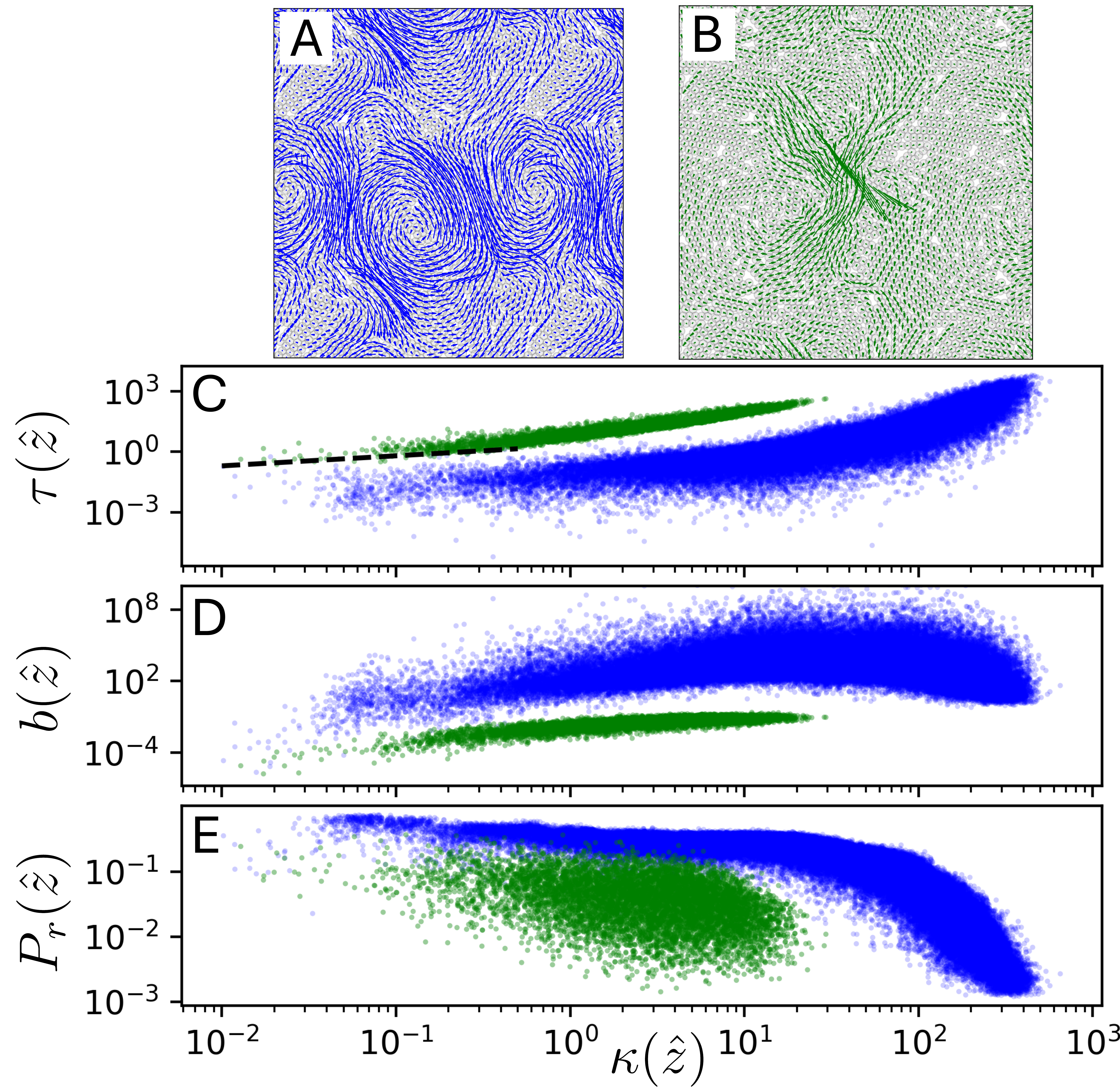}
\caption{A) An example low-$\kappa$ harmonic mode measured in a non-Hamiltonian, active solid. B) An example low-$b$ cubic mode from the same configuration as A). Note that while these modes consist of a translational and rotational component for each rod, we visualize them by plotting the resulting translations of the spherical sub-particles that comprise the rods. Additional details in the Supplemental Information. C-E) Distributions of the asymmetry $\tau(\hat{z})$ (C), energy barrier $b(\hat{z})$ (D), and participation ratio $P_r(\hat{z})$ (E) of an ensemble of harmonic modes (blue) and cubic modes (green) as a function of stiffness $\kappa(\hat{z})$. \tah{Dashed line in (C) indicates a scaling of $\tau \propto \sqrt{\kappa}$.}}
\label{CubicVsHarmonic}
\end{figure}

Fig \ref{ForceLandscapeCartoon_StressStrain}E shows a strain-stress curve obtained by quasistatically increasing $\sigma_\text{act}$, where plastic instabilities are clearly visible as discontinuous jumps in strain.  We characterize the plastic deformation that occurs during a strain jump as the magnitude of non-affine displacements $D^2_{\text{min}}$~\cite{falk_dynamics_1998} around each sub-particle. As a single instability typically consists of an avalanche of discrete plastic events~\cite{stanifer_avalanche_2022}, we cluster the $D^2_{\text{min}}$ field into separate rearrangements using persistent homology, which can robustly separate signal from noise in such systems \cite{stanifer_avalanche_2022} (see Supplemental Materials). In Fig~\ref{Clustering+Results}C,D, respectively, we show an example $D^2_{\text{min}}$ field and the resulting two clusters of particles that underwent rearrangements.

At each value of $\sigma_\text{act}$ we numerically obtain populations of harmonic and cubic modes, see Supplemental Materials. Fig~\ref{CubicVsHarmonic}(A) shows a low-stiffness, phonon-like harmonic mode and a quasilocalized cubic mode (B) with a disordered core and quadrupolar decaying field, as in non-active amorphous systems. We use persistent homology to obtain a cluster of particles in the localized core of each cubic mode, and quantify the correlation between low-$b(\hat{\pi})$ cubic modes and upcoming plastic events with a metric called proficiency (denoted by $\chi$), see Supplemental Materials. This is a normalized measure (ranging from 0 to 1) of the mutual information between a given cluster of rearranging particles and a previously predicted soft spot \cite{stanifer_avalanche_2022}.

\noindent
\textbf{Results.}
Cubic modes generally have larger $\kappa$ than the lowest-stiffness harmonic modes~\cite{gartner_nonlinear_2016}, and Fig~\ref{CubicVsHarmonic}C shows that active cubic modes also have much larger $\tau$. \tah{Passive spheres obey the scaling relation $\tau\!\propto\!\sqrt{\kappa}$~\cite{kapteijns_nonlinear_2020} (dashed line in Fig \ref{CubicVsHarmonic}C). Our system follows this scaling for small $\kappa$, switching over to $\tau \propto \kappa$ at larger $\kappa$.} As expected, the larger $\tau$ generates lower energy barriers $b$ for cubic modes compared to harmonic modes (Fig \ref{CubicVsHarmonic}D). Cubic modes also have significantly smaller participation ratios $P_r(\hat{z})\! =\! \langle\hat{z}_i^2\rangle^2/\langle\hat{z}_i^4\rangle$ than the low-stiffness harmonic modes (Fig \ref{CubicVsHarmonic}e), implying that they are more localized.

In Fig \ref{Clustering+Results}E-G, we show data from a single simulation of the six lowest-lying cubic and harmonic modes as the system is driven through several instabilities, indicated by vertical black lines. Similar to non-active amorphous solids under quasistatic shear \cite{maloney_amorphous_2006, manning_vibrational_2011}, we see that the onset of an instability is preceded by a single harmonic mode approaching zero stiffness (Fig \ref{Clustering+Results}E), as just before a bifurcation the harmonic approximation exactly identifies mode that will become unstable \cite{maloney_amorphous_2006, lerner_micromechanics_2016}. Further from an instability, the lowest-stiffness harmonic mode quickly becomes decorrelated with the unstable mode $\hat{\phi}_{\text{us}}$. In Fig \ref{Clustering+Results}E each point is colored by the overlap $\hat{\phi}\cdot \hat{\phi}_{\text{us}}$ between the corresponding harmonic mode $\hat{\phi}$ and the unstable mode $\hat{\phi}_{\text{us}}$ that appears at the last instability at  $\sigma_{\text{act}} \approx 0.00088$.

In Fig \ref{Clustering+Results}F we plot the stiffness of the six lowest cubic modes colored by their overlaps $\hat{\pi}\cdot \hat{\phi}_{\text{us}}$ with this unstable mode. Similarly, the onset of instabilities are preceded by a single cubic mode approaching zero stiffness. In contrast to harmonic modes, there is a cubic mode that maintains high overlap with the unstable mode over a wide interval before the instability, even multiple avalanches in advance. This suggests that cubic modes might retain predictive power over many plastic events.

The mode that becomes unstable to initiate an avalanche does not predict the entire plastic deformation that occurs during the avalanche, as it often triggers additional defects. Our persistent homology clustering allows us to identify individual rearrangements that comprise the avalanche and order them in time and strain. In Fig \ref{Clustering+Results}A-B we show the two cubic modes with the lowest energy barriers and their corresponding soft spots calculated slightly before the avalanche shown in Fig \ref{Clustering+Results}C. These two soft spots are correlated with the two rearrangements in the upcoming avalanche shown in Fig \ref{Clustering+Results}D (yellow, which occurs first and red which occurs second), and have correspondingly high values of proficiency ($\chi=0.17, 0.35$), while a typical random cluster has $\chi \lessapprox 10^{-2}$, see Supplemental Materials. In Fig \ref{Clustering+Results}G we plot the six cubic modes with the lowest energy barriers colored by their proficiency with the first rearrangement of the avalanche at $\sigma_{\text{act}} \approx 0.00088$. As expected, the cubic mode that highly overlaps the unstable mode also has a high proficiency with the first rearrangement.

\begin{figure}[h!]
\centering
\includegraphics[scale= 0.10]{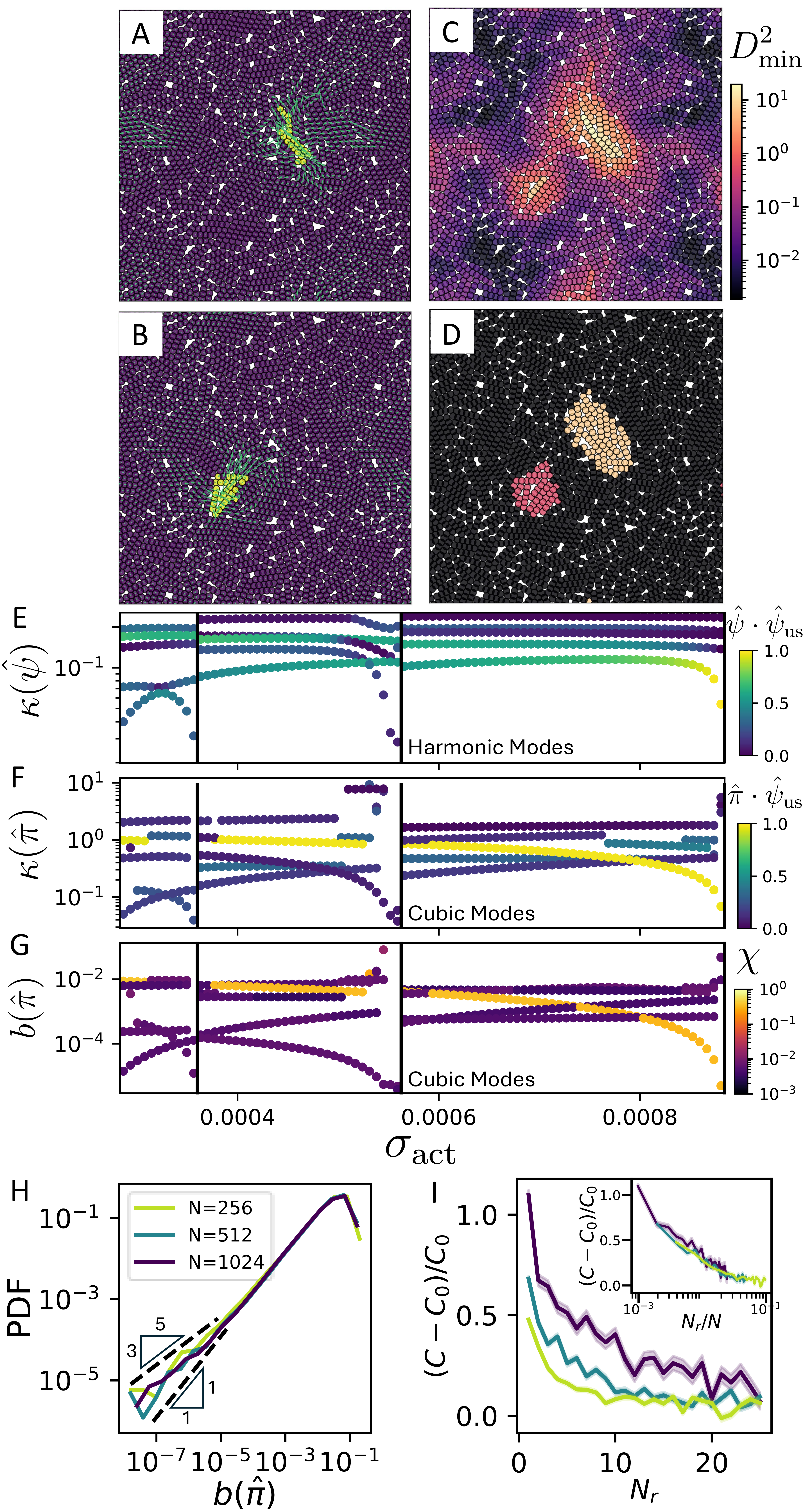}
\caption{A,B) The two cubic modes with the lowest $b(\hat{\pi})$ observed $\Delta\sigma_{\text{act}} = 10^{-4}$ before the upcoming avalanche shown in C), with corresponding soft spots highlighted. C) A $D^2_{\text{min}}$ field showing the plastic deformation during the avalanche. D) Two clusters of particles obtained through persistent homology corresponding with localized plastic events that comprise the avalanche in C). Note that the $D^2_{\text{min}}$ field and clustering are calculated with the individual spherical sub-particles that comprise the rods. E,F) $\kappa$ values over a range of $\sigma_{\text{act}}$ for the 6 lowest harmonic modes (E) and cubic modes (F) colored by their overlaps with the unstable mode $\hat{\phi}_{\text{us}}$ at $\sigma_{\text{act}} \approx 0.00088$. G) $b(\hat{\pi})$ values for the 6 lowest cubic modes, colored by their proficiency with the rearrangement at $\sigma_{\text{act}} \approx 0.00088$. \tah{H) Distribution of cubic mode energy barriers. Dashed lines indicate power laws $P(b) \propto b^{0.6}$ and $P(b) \propto b$. I) Excess fraction of successful predictions (relative to random clusters) of upcoming rearrangements. Inset shows same data rescaled by the system size $N$ to demonstrate collapse.}}
\label{Clustering+Results}
\end{figure}

How far in advance of a rearrangement is there a predictive cubic mode with a high proficiency? To quantify this, we measure the probability $C(N_m, N_r)$ that one of the $N_m$ lowest-$b$ cubic modes of a configuration has a high proficiency (greater than a threshold value $\chi_\text{th}=0.02$, see Supplemental) with the $N_r$'th upcoming rearrangement. We must limit the number of modes $N_m$ because our protocol generates upwards of $100$ cubic modes per packing that cover nearly the entire system and guarantee high overlap just by chance, and we are able to choose $N_m(N)$  in a principled way by comparing to a baseline probability $C_0(N_m)$ for Poisson-distributed random clusters, see Supplemental Materials.

In Fig \ref{Clustering+Results}I we plot the relative increase of successful correlations over random clusters $(C - C_0)/C_0$. For the smallest system size ($N=256$) cubic modes are about 50\% better than chance at predicting the next rearrangement, but this decreases until about the 10th rearrangement which shows no improvement. However, in the largest system size ($N=1024$) cubic modes are over twice as good at predicting the first rearrangement, with a slight benefit over random clusters still at the 20th rearrangement. This makes sense as larger systems require more rearrangements before their structure becomes decorrelated. In the inset to Fig \ref{Clustering+Results}(I), we plot the same data rescaled by the system size, which collapses onto a single curve.  This confirms that force-landscape-based nonlinear modes successfully predict plasticity, and indeed are predictive over a substantially larger number of rearrangements than vibrational modes (one of the best ways to predict plasticity~\cite{richard_predicting_2020}) in non-active spherical particle systems~\cite{stanifer_avalanche_2022}.

We can also extract statistical properties of defects. Fig \ref{Clustering+Results}(H) shows the statistics of energy barriers for cubic modes in our system, which exhibit power law statistics for low barrier heights. In passive particulate glasses, it has been suggested that the probability distribution of energy barriers should scale universally as $b^{1/4}$~\cite{kapteijns_nonlinear_2020}. Here, we find an exponent between $3/5$ and $1$; future work could focus on pinpointing this exponent and understanding how it changes as a function of activity, material preparation, and particle anisotropy.

\noindent
\textbf{Discussion.} We have identified defects in dense non-Hamiltonian active solids by computing the analogue of harmonic and nonlinear excitations around stable points of a composite `force landscape'. Harmonic excitations are sensitive to imminent instabilities, while nonlinear excitations are highly correlated with plastic rearrangements far in advance, over 10-20 rearrangements in small systems. 

Since our definition of defects explicitly includes active forces, extensions of this work could use it as a starting point for control. We focus here on systems that were initially force-balanced with a well-defined reference state, and then driven to an instability by increasing the active forces on all particles uniformly.  Instead, one could drive the system externally (such as with a global shear strain) while allowing the active force of each particle to vary individually as a 'tuning parameter' controlled internally or externally~\cite{reyes_garza_magnetic_2023}. Since our method allows us to compute how changes to active force alter the defects, one could in principle use it to develop a tuning rule~\cite{stern_learning_2023, dillavou_demonstration_2022} to activate flow, resist flow, or change the effective viscosity in different locations within the material.

There are reasons to expect this approach may also work in slowly flowing systems without a well-defined reference state. In passive athermal systems, extensions of excitation-based landscape approaches accurately predict the dynamics even when the system is moving downhill in the energy landscape~\cite{stanifer_avalanche_2022}. Similarly, dynamical heterogeneities in supercooled fluids have been shown to behave like avalanches composed of localized excitations dominated by the underlying landscape~\cite{tahaei_scaling_2023}.

This approach can also constrain and extend constitutive laws for dense active matter.  Elastoplastic and fluidity models qualitatively capture macroscopic rheology of dense non-active~\cite{nicolas_deformation_2018, barlow_ductile_2020} and active/non-reciprocal matter~\cite{ghosh_elastoplastic_2025, ioratim-uba_exotic_2025, ghaznavi_yielding_2025}. To make such theories quantitative, one needs to specify the distribution $P(x)$ of 'strain required for a defect to yield' ($x$). In passive glasses~\cite{lerner_micromechanics_2016} and supercooled fluids~\cite{lerbinger_relevance_2022}, the statistics of nonlinear energy barriers $b(\hat{z})$ -- Fig~\ref{Clustering+Results}(H) -- have been directly related to  $P(x)\sim x^{\theta}$, with $\theta \sim 0.6$~\cite{karmakar_statistical_2010, lin_density_2014, barbot_local_2018}, via another scaling relation $x \sim \kappa^2 / \tau$~\cite{lerner_micromechanics_2016, kapteijns_nonlinear_2020}. It would be interesting to see whether the latter scaling relation still holds in active systems, and then extract $P(x)$. In some systems, there is evidence for dynamics beyond standard elastoplastic models and include structural interactions between defects~\cite{ghaznavi_yielding_2025} via, e.g. 'Structural Elastoplastic' (StEP) models~\cite{zhang_structuro-elasto-plasticity_2022, xiao_identifying_2023}, which could also be quantified via this approach. 

Beyond systems with non-conservative self-propulsion forces and conservative contact forces, our formalism extends to systems with nonreciprocal interactions or any differentiable force law. This should enable prediction and control over a large class of dense matter where forces do not arise via potentials.

\noindent
\textbf{Acknowledgments} TH and MLM were primarily supported by DMR-CMMT-2532170 and DMR-CMMT-1951921. MLM also acknowledges support from grant number 2023‐329572 from the Chan Zuckerberg Initiative DAF, an advised fund of Silicon Valley Community Foundation.  

\bibliographystyle{unsrt}
\bibliography{ActiveRodsBib}

% \clearpage
\clearpage

\section{Supplemental Materials}

\setcounter{figure}{0}
\renewcommand{\figurename}{Fig.}
\renewcommand{\thefigure}{S\arabic{figure}}

\subsection{Defining linear and nonlinear excitations in a force landscape}
\subsubsection{Linear and nonlinear excitations in a potential energy landscape}
Let us consider a system with degrees of freedom $\{x_i\}$ and an energy functional $U(\{x_i\})$. Assuming the system is in mechanical equilibrium we can write the energy cost of a perturbation $\delta x_i = s\hat{z}_i$ up to 3rd order in $s$ as
\begin{align}
    \delta U &= \frac{1}{2}\left[ \sum_{ij} \frac{\partial^2 U}{\partial x_i \partial x_j}\hat{z}_i \hat{z}_j \right] s^2 \\
    &+ \frac{1}{6}\left[ \sum_{ijk} \frac{\partial^2 U}{\partial x_i \partial x_j \partial x_k}\hat{z}_i \hat{z}_j \hat{z}_k \right] s^3 + \mathcal{O}(s^4), \\
    &= \frac{1}{2}\kappa(\mathbf{\hat{z}})s^2 + \frac{1}{6}\tau(\mathbf{\hat{z}}) s^3 + \mathcal{O}(s^4),
    \label{energy expansion}
\end{align}
where $\mathbf{\hat{z}}$ is a unit vector in phase space denoting the direction of the perturbation and $s$ is the amplitude of the displacement. The coefficients $\kappa(\mathbf{\hat{z}})$ and $\tau(\mathbf{\hat{z}})$ are given by
\begin{align}
    \kappa(\mathbf{\hat{z}}) &= \sum_{ij} \frac{\partial^2 U}{\partial x_i \partial x_j}\hat{z}_i \hat{z}_j = \sum_{ij} \mathcal{M}_{ij}\hat{z}_i \hat{z}_j \label{stiffness definition},\\
    \tau(\mathbf{\hat{z}}) &= \sum_{ijk} \frac{\partial^2 U}{\partial x_i \partial x_j \partial x_k}\hat{z}_i \hat{z}_j \hat{z}_k = \sum_{ijk} \mathcal{T}_{ijk}\hat{z}_i \hat{z}_j \hat{z}_k, \label{asymmetry definition}
\end{align}
where we define the tensors of second- and third-order derivatives of the energy $\mathcal{M}$ and $\mathcal{T}$. As $\mathcal{M}$ is simply the Hessian matrix, $\kappa(\mathbf{\hat{z}})$ describes the curvature of the energy landscape along the direction $\mathbf{\hat{z}}$, while $\tau(\mathbf{\hat{z}})$ gives a measure of the asymmetry of the energy landscape in this direction.

With this cubic approximation of the energy landscape we can write the quantity 
\begin{align}
    b(\mathbf{\hat{z}}) = \frac{2}{3}\frac{\kappa(\mathbf{\hat{z}})^3}{\tau(\mathbf{\hat{z}})^2},
    \label{barrier definition}
\end{align}
which is the energy difference between the local maximum and minimum of the cubic approximation to the energy landscape $\delta U_{\text{cubic}}=1/2\,\kappa(\mathbf{\hat{z}})s^2 + 1/6\,\tau(\mathbf{\hat{z}}) s^3$ along the direction $\mathbf{\hat{z}}$. While the expression in Eqn. \ref{barrier definition} is not an exact description of the energy barriers, displacement modes that minimize Eqn. \ref{barrier definition} (known as nonlinear plastic modes) have been shown to be the quasilocalized, and any un-hybridized quasilocalized modes that do appear in the harmonic spectrum are very nearly minima of Eq. \ref{barrier definition}. Previous work has used gradient descent to minimize Eqn. \ref{barrier definition} in order to generate populations of nonlinear plastic modes from a set of initial guesses derived from the linear response to force dipoles.

\subsubsection{Nonlinear excitations in a force landscape}

A similar analysis can be performed without making explicit reference to an energy function. Instead of expanding the energy cost of a perturbation, we can directly expand the force response of the system nearby a force-balanced reference state as
\begin{align}
    \mathbf{f}(\mathbf{\hat{z}}, s) = \mathbf{f}^{(1)}(\mathbf{\hat{z}})s + \frac{1}{2}\mathbf{f}^{(2)}(\mathbf{\hat{z}})s^2 + \mathcal{O}(s^3). \label{force expansion}
\end{align}
First, let us assume the force is derived from an energy function ($\mathbf{f} = -\nabla U$), in which case the coefficients $\mathbf{f}^{(1)}(\mathbf{\hat{z}})$ and $\mathbf{f}^{(2)}(\mathbf{\hat{z}})$ in the expansion can be written in terms of the derivatives of the energy as
\begin{align}
    f^{(1)}_i(\mathbf{\hat{z}}) &= \sum_{j} \frac{\partial f_i}{\partial x_j}\hat{z}_j = -\sum_{j} \mathcal{M}_{ij}\hat{z}_j \label{f1 coefficient from energy} \\
    f^{(2)}_i(\mathbf{\hat{z}}) &= \sum_{jk} \frac{\partial^2 f_i}{\partial x_j \partial x_k}\hat{z}_j \hat{z}_k = -\sum_{jk} \mathcal{T}_{ijk}\hat{z}_j \hat{z}_k \label{f2 coefficient from energy}
\end{align}

Relationships between these coefficients can be used to define special populations of excitations. For example, harmonic modes are defined as displacements such that the resulting first-order force response is parallel to the imposed displacements:
\begin{align}
    \mathbf{f}^{(1)}(\mathbf{\hat{z}}) \propto \mathbf{\hat{z}} \label{harmonic mode definition}
\end{align}
For systems with conservative forces, these harmonic modes are exactly the vibrational normal modes, since Eq \ref{harmonic mode definition} is satisfied whenever $\hat{z}$ is an eigenvector of the Hessian with some eigenvalue $\kappa$:
\begin{align}
    f^{(1)}_i(\mathbf{\hat{z}}) = -\sum_{j} \mathcal{M}_{ij}\hat{z}_j = -\kappa \hat{z}_i \propto \hat{z}_i \label{harmonic mode definition 2}
\end{align}

We extend this concept to a family of nonlinear excitations by defining an $n$'th order nonlinear mode as a solution to 
\begin{align}
    \mathbf{f}^{(n-1)}(\mathbf{\hat{z}}) \propto \mathbf{f}^{(1)}(\mathbf{\hat{z}}), \label{general nonlinear mode definition}
\end{align}
where $\mathbf{f}^{(n-1)}(\mathbf{\hat{z}})$ is the $(n-1)$'th order coefficient of the force expansion in Eq \ref{force expansion} (which corresponds with the $n$'th order term in the expansion of the energy in Eq \ref{energy expansion}). In this work we will focus on 3rd order nonlinear modes (which we will refer to as cubic modes) that satisfy
\begin{align}
    \mathbf{f}^{(2)}_i(\mathbf{\hat{z}}) \propto \mathbf{f}^{(1)}_i(\mathbf{\hat{z}}). \label{cubic mode definition 2}
\end{align}
We can then substitute Eqs. \ref{f1 coefficient from energy},\ref{f2 coefficient from energy} and contract with $\mathbf{\hat{z}}$ to obtain the proportionality constant and write an equality condition that cubic modes must satisfy:
\begin{align}
    \sum_{jk} \mathcal{T}_{ijk}\hat{z}_j \hat{z}_k &= c \sum_{j} \mathcal{M}_{ij}\hat{z}_j\\
    \tau(\mathbf{\hat{z}}) &= c \,\kappa(\mathbf{\hat{z}}) \\
    \tau(\mathbf{\hat{z}}) \sum_{j} \mathcal{M}_{ij}\hat{z}_j  &- \kappa(\mathbf{\hat{z}}) \sum_{jk} \mathcal{T}_{ijk}\hat{z}_j \hat{z}_k = 0. \label{cubic mode definition 3}
\end{align}
With this expression, we can show that the nonlinear plastic modes found by minimizing $b(\mathbf{\hat{z}})$ are also cubic modes. In fact, any critical point of $b(\mathbf{\hat{z}})$ is a cubic mode, since
\begin{align}
    \frac{\partial b}{\partial \hat{z}_i} &= \frac{2}{3}\left(3 \frac{\kappa^2}{\tau^2} \frac{\partial \kappa}{\partial \hat{z}_i} -2 \frac{\kappa^3}{\tau^3} \frac{\partial \tau}{\partial \hat{z}_i} \right)\label{barrier grad 1}\\
    &= 4\frac{\kappa^2}{\tau^3}\left(\frac{\tau}{2} \frac{\partial \kappa}{\partial \hat{z}_i} - \frac{\kappa}{3} \frac{\partial \tau}{\partial \hat{z}_i} \right)\label{barrier grad 2}\\
    &= 4\frac{\kappa^2}{\tau^3}\left(\tau \sum_{j} \mathcal{M}_{ij}\hat{z}_j  - \kappa \sum_{jk} \mathcal{T}_{ijk}\hat{z}_j \hat{z}_k \right),\label{barrier grad 3}
\end{align}
which vanishes exactly when Eq. \ref{cubic mode definition 3} is satisfied.

\subsubsection{Including Non-Conservative Forces}
Now we would like to extend this formalism to systems with non-conservative force laws:
\begin{align}
    \mathbf{f}(\mathbf{x}) = -\nabla U + \mathbf{f}^{\text{nc}}(\mathbf{x}), \label{active force}
\end{align}
where we include a separate terms for conservative forces $-\nabla U$ (such as those arising from a pair-wise contact potential) and non-conservative forces $\mathbf{f}^{\text{nc}}$ which could come from active self-propulsion forces or from non-reciprocal interactions. We assume that these non-conservative forces depend directly on the current state of the system $\mathbf{x}$, and that any other control parameters (e.g. magnitudes of self-propulsion forces or strength of interactions) vary slowly; that is, on time scales larger than the relaxation time of the system.

Now, the coefficients in the force expansion in Eq. \ref{force expansion} are given by 
\begin{align}
    f^{(1)}_i(\mathbf{\hat{z}}) &= -\sum_{j}\left(\frac{\partial^2 U}{\partial x_i \partial x_j} - \frac{\partial f_i^{\text{nc}}}{\partial x_j}\right) \hat{z}_j = -\sum_{j} \tilde{\mathcal{M}}_{ij}\hat{z}_j \label{f1 coefficient from forces} \\
    f^{(2)}_i(\mathbf{\hat{z}}) &= -\sum_{jk} \left(\frac{\partial^3 U}{\partial x_i \partial x_j \partial x_k} - \frac{\partial^2 f_i^{\text{nc}}}{\partial x_j \partial x_k}\right) \hat{z}_j \hat{z}_k \nonumber\\
    &= -\sum_{jk} \tilde{\mathcal{T}}_{ijk}\hat{z}_j \hat{z}_k. \label{f2 coefficient from forces}
\end{align}
These expressions are very similar to Eqs \ref{f1 coefficient from energy} and \ref{f2 coefficient from energy}, except that the tensors of derivatives $\mathcal{M}$ and $\mathcal{T}$ have been replaced by the ``augmented" tensors $\tilde{\mathcal{M}}$ and $\tilde{\mathcal{T}}$ that also account for how the non-conservative forces change as the system is perturbed. Note that the non-conservative nature of the forces implies that the augmented derivative tensors are no longer necessarily symmetric. The code in our open source repository \cite{active-rods_GitHub} provides a detailed calculations of $\tilde{\mathcal{M}}$ and $\tilde{\mathcal{T}}$ for our system of self-propelled rods.

Since in Eqs \ref{harmonic mode definition 2} and \ref{cubic mode definition 2} we defined harmonic and cubic modes solely in terms of these force expansion coefficients, the generalization to non-conservative forces involves replacing the derivative tensors in the above expressions with their augmented versions. From Eq \ref{harmonic mode definition 2}, harmonic modes are now eigenvectors of the augmented Hessian:
\begin{align}
    f^{(1)}_i(\mathbf{\hat{z}}) = -\sum_{j} \tilde{\mathcal{M}}_{ij}\hat{z}_j = -\kappa \hat{z}_i \label{harmonic mode definition nc}
\end{align}
From Eq \ref{cubic mode definition 3}, cubic modes satisfy
\begin{align}
    \tau(\mathbf{\hat{z}}) \sum_{j} \tilde{\mathcal{M}}_{ij}\hat{z}_j  &- \kappa(\mathbf{\hat{z}}) \sum_{jk} \tilde{\mathcal{T}}_{ijk}\hat{z}_j \hat{z}_k = 0, \label{cubic mode definition nc}
\end{align}
where
\begin{align}
    \kappa(\mathbf{\hat{z}}) &= \sum_{ij} \tilde{\mathcal{M}}_{ij}\hat{z}_i \hat{z}_j \label{stiffness definition nc}\\
    \tau(\mathbf{\hat{z}}) &= \sum_{ijk} \tilde{\mathcal{T}}_{ijk}\hat{z}_i \hat{z}_j \hat{z}_k, \label{asymmetry definition nc}
\end{align}
Note that even without an energy landscape, $\kappa(\mathbf{\hat{z}})$, $\tau(\mathbf{\hat{z}})$, and $b(\mathbf{\hat{z}}) = 2/3 \kappa^2/\tau^3$ can still be interpreted as a stiffness, asymmetry, and energy barrier respectively. We can write the cubic approximation for the work required to deform the system a magnitude $s$ along a direction $\mathbf{\hat{z}}$ as
\begin{align}
    &W(\mathbf{\hat{z}}, s) = -\int_0^s \text{d}s \;\mathbf{\hat{z}}\cdot \mathbf{f}(\mathbf{\hat{z}}, s) \\
    &= \int_0^s \text{d}s \;\mathbf{\hat{z}}\cdot \left( \mathbf{f}^{(1)}(\mathbf{\hat{z}})s + \frac{1}{2}\mathbf{f}^{(2)}(\mathbf{\hat{z}})s^2 \right) + \mathcal{O}(s^4)\\
    &= \frac{1}{2}\kappa(\mathbf{\hat{z}}) s^2 + \frac{1}{6}\tau(\mathbf{\hat{z}}) s^3 + \mathcal{O}(s^4). \label{work}
\end{align}
Then $\kappa(\mathbf{\hat{z}})$ and $\tau(\mathbf{\hat{z}})$ still represent the stiffness and asymmetry that an external driving force would feel, and $b(\mathbf{\hat{z}}) = 2/3 (\kappa^3/\tau^2)$ is (a cubic approximation of) the amount of work that must be done before the internal reaction forces of the system begin acting in favor of the imposed deformation $\mathbf{\hat{z}}$ rather than driving the system back to its stable reference configuration.

Note that this does not imply that there is a global energy landscape. Eq \ref{work} gives the work required to deform a system along a fixed direction, but in non-equilibrium systems this is generally path-dependent. For example, it is possible to extract energy from active systems as they traverse a closed loop in phase space. So Eq \ref{work} can be thought of as describing a collection of 'work' landscapes along different directions, but these cannot be combined into a single global landscape.

A consequence of this is that minimizing the ``energy barrier" $b(\mathbf{\hat{z}})$ no longer produces cubic modes. The equality between Eqns \ref{barrier grad 2} and \ref{barrier grad 3} does not generically hold if there are non-conservative forces. In general 
\begin{align}
    \frac{\partial \kappa}{\partial \hat{z}_i} = \sum_j\tilde{\mathcal{M}}_{ij}\hat{z}_j + \tilde{\mathcal{M}}_{ji}\hat{z}_j,
\end{align}
so that cubic modes which satisfy Eqn \ref{cubic mode definition nc} are not critical points of $b(\mathbf{\hat{z}})$ unless $\tilde{\mathcal{M}}$ and $\tilde{\mathcal{T}}$ are symmetric. Our attempts to numerically minimize $b(\mathbf{\hat{z}})$ in our system of self-propelled rods by gradient descent generally did not converge, which we attribute to the lack of a global energy landscape due to the non-conservative forces.

Instead, we numerically generate cubic modes by using the fact that solutions to Eq \ref{cubic mode definition nc} are fixed points of the discrete mapping
\begin{align}
    \hat{z}^{(n+1)}_i = \frac{\kappa(\mathbf{\hat{z}}^{(n)})}{\tau(\mathbf{\hat{z}}^{(n)})}\sum_{jkl} \tilde{\mathcal{M}}^{-1}_{ij}\tilde{\mathcal{T}}_{jkl}\hat{z}^{(n)}_k\hat{z}^{(n)}_l, \label{cubic mode mapping}
\end{align}
where $n$ indicates the iteration number. As this does not rely on any symmetries of the tensors $\tilde{\mathcal{M}}$ and $\tilde{\mathcal{T}}$, we find (similar to previous studies in non-active systems \cite{lerner_micromechanics_2016, gartner_nonlinear_2016-1}) that under this mapping nearly all initial conditions $\mathbf{\hat{z}}^{(0)}$ converge to a cubic mode that satisfies Eqn \ref{cubic mode definition nc} within numerical precision in around 10-30 iterations (although rarely it will instead converge to a limit cycle).

\subsection{Details of Numerical Methods}
\subsubsection{Simulation interactions and dynamics}
\begin{figure}[h]
\centering
\includegraphics[scale= 1]{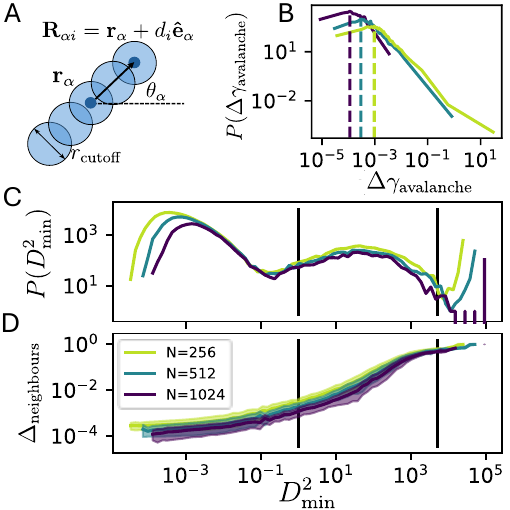}
\caption{A) Illustration of the various quantities associated with a single rod and its 5 constituent particles. B) Distribution of strain intervals $\Delta\gamma_{\text{avalanche}}$ between the onset of subsequent instabilities for different system sizes (colors in all plots given by legend in D). The most likely strain for each system size is denoted by a dashed line. C) Distribution of the maximum value of $D^2_{\text{min}}$ between subsequent stable configurations. D) Average fraction of particles that experience a neighbor change $\Delta_{\text{neighbors}}$ at different amounts of non-affine displacement (shaded regions show one standard deviation from the mean). Black lines indicate upper and lower limits on avalanches that we consider: that is, only $D^2_{\text{min}}$ fields with maximum values between these lines is clustered for further analysis.}
\label{FigSM1}
\end{figure}

Following previous implementations of self-propelled rods~\cite{wensink_meso-scale_2012, grauer_spontaneous_2018, janssen_aging_2017}, we model each rod as a set of 5 spherical particles to simplify the calculation of interaction forces. The degrees of freedom are the center of mass coordinates $\mathbf{r}_\alpha$ and the angles $\theta_\alpha$ between the ``forward" directions of the rods and the $x-$axis, where $\alpha=1,...,N$ indexes the rods. These determine the coordinates of the sub-particles
\begin{align}
    \mathbf{R}_{\alpha i} = \mathbf{r}_\alpha + d_i \mathbf{\hat{e}}_\alpha,
\end{align}
where $i=1,...,5$ indexes the particles in a single rod, $\mathbf{\hat{e}}_\alpha = [\text{cos}\,\theta_\alpha \;\, \text{sin}\,\theta_\alpha]^T$ is the unit vector along the forward direction of rod $\alpha$, and $d_i = 0,1,-1,2,-2$ is the displacement of each particle along $\mathbf{\hat{e}}_\alpha$ relative to the center of mass, which sets the length scale of the system. These quantities are illustrated in Fig \ref{FigSM1}A.

The interaction energy is given by 
\begin{align}
    U = \sum_{\alpha<\beta}\sum_{ij}u(|\mathbf{R}_{\alpha i} - \mathbf{R}_{\beta j}|),
\end{align}
where $u(r)$ is a pair-wise repulsive interaction between particles of different rods. In this study we use an inverse power law potential
\begin{align}
    u(r) =\begin{cases} 
      \epsilon(r^{-10} + c_6 r^6 + c_4 r^4 + c_2 r^2 + c_0) & r<r_{\text{cutoff}} \\
      0 & r\geq r_{\text{cutoff}}
   \end{cases}
\end{align}
where $\epsilon$ is an energy scale which we set to $1$ and the coefficients are chosen such that $u$ and its first three derivatives are continuous at the cutoff distance $r_{\text{cutoff}}=1.48$. This cutoff is chosen to be slightly larger than the separation between adjacent particles in a rod in order to decrease the apparent ``bumpiness" of the rods. The net forces on the particles $\mathbf{F}_{\alpha i}$ are then given by
\begin{align}
 \mathbf{F}_{\alpha i} = \sum_{\beta\neq\alpha, j} -u'(|\mathbf{R}_{\alpha i} - \mathbf{R}_{\beta j}|) \frac{\mathbf{R}_{\alpha i} - \mathbf{R}_{\beta j}}{|\mathbf{R}_{\alpha i} - \mathbf{R}_{\beta j}|},
\end{align}
which can be summed with the active forces $\mathbf{f}_{\alpha}^{\text{nc}} = \eta \mathbf{\hat{e}}_\alpha$ to obtain the net forces $\mathbf{f}_{\alpha}$ and torques $T_\alpha$ on the rods:
\begin{align}
    \mathbf{f}_{\alpha} &= \sum_i \mathbf{F}_{\alpha i} + \eta\mathbf{\hat{e}}_\alpha - \eta\langle\mathbf{\hat{e}}_\beta\rangle_\beta\\
    T_{\alpha} &= \sum_{i} d_i\mathbf{\hat{e}}_\alpha \times \mathbf{F}_{\alpha i}, \label{active rods forces}
\end{align}
where $\eta$ is the magnitude of the active force on a single rod. Note that we also subtract the average of the active forces $\eta\langle\mathbf{\hat{e}}_\beta\rangle_\beta$, which averages to zero in the thermodynamic limit but scales as $\sqrt{N}$ for finite systems. This has the effect of eliminating any overall drift velocity and fixing the center of mass of the system, but it is also important to include in the calculation of the tensors $\tilde{\mathcal{M}}$ and $\tilde{\mathcal{T}}$, as the derivations above assume an expansion around a state with no net forces.

We then evolve the system with overdamped dynamics according to the equations of motion
\begin{align}
    \frac{\partial \mathbf{r}_\alpha}{\partial t} &= [D_{\parallel}(\mathbf{\hat{e}}_\alpha \mathbf{\hat{e}}_\alpha^T) + D_{\perp}(\mathbf{1} - \mathbf{\hat{e}}_\alpha \mathbf{\hat{e}}_\alpha^T)]\mathbf{f}_{\alpha} \label{equation of motion}\\
    \frac{\partial \theta_{\alpha}}{\partial t} &= D_{\text{rot}} T_{\alpha},
\end{align}
which in general includes different diffusion constants for rotation ($D_{\text{rot}}$) and translational motion directed either parallel ($D_\parallel$) or perpendicular ($D_\perp$) to the long direction of a rod due to the geometric dependence of drag forces. However, as our study does not directly concern these effects, for simplicity we set all three diffusion constants numerically to 1 to have a single time scale for the dynamics. We do not expect this choice to significantly affect our results, especially as we are focused on a regime where the system is almost fully arrested.

We initialize the system at a low number density ($\rho=5N/L^2=0.2$) in a square periodic domain, place the rod's centers of mass on a square grid and assign rod angles sampled from a uniform distribution. We then evolve the system with a high level of activity ($\eta=1$) to reach a mixed steady state, then turn off the activity ($\eta=0$) and quasistatically reduce the size of the periodic domain until the target number density is reached ($\rho=0.7, 0.75$). These densities are substantially higher than the jamming transition, which occurs around $\rho=0.6$ in this system, as the slow dynamics close to unjamming cause relaxation times to become very long and the length scale of particle rearrangements approaches the system size. At these high densities, we did not observe substantial dependence of defects or plasticity on $\rho$, and so we report only data from systems with higher density as they remain in an arrested state for a wider range of activity and thus produce better statistics.

The system was driven through instabilities by quasistatically increasing the \mlm{average magnitude of the active force field. We started from zero magnitude and linearly increased the magnitude up to a system-dependent threshold at which the system became fluidized, e.g. no longer has a well-defined reference state. We defined the fluidized state as a system with a diverging avalanche sizes, to the right of the right-most vertical line in Fig.~\ref{FigSM1}(C,D). This protocol ensured that we investigated the full range of activity magnitudes that still permitted a well-defined reference state.  If we had driven the system with an external strain, the activity may have acted only as an insignificant perturbation. Instead, the 'ramping magnitude of force protocol' we applied here provides a strong test of our method even in the regime where the active forces are comparable to the interparticle interactions. We note that a similar protocol was recently used to study the jamming behavior of active spheres~\cite{gandikota_jammed_2026}.}

\mlm{One downside to this protocol is that leads us to consider together (typically small) avalanches that occur far from the macroscopic material yielding transition~\cite{berthier_yielding_2025} and (typically larger) avalanches that occur close to the macroscopic yielding transition. The fact that we do not treat these regimes differently likely decreases our ability to predict rearrangements.}

To quantify the response to the activity we follow \cite{morse_direct_2021} and write the stress due to the field of active forces as $\sigma_{\text{act}} = \eta/L\sqrt{12}$, and $L$ is the linear system dimension. This normalization is such that the associated conjugate strain $\gamma_{\text{act}}$ can be directly compared with that of a global shear. The strain is proportional to the mean displacement of the rods along the directions of their active forces, which is numerically integrated during the time evolution:
\begin{align}
    \gamma_{\text{act}} = \frac{\sqrt{12}}{NL} \int_{t_0}^{t_1} \text{d}t \sum_\alpha \frac{\partial \mathbf{r}_\alpha}{\partial t} \cdot \mathbf{\hat{e}}_\alpha.
\end{align}

We increase the active stress in steps of $\delta\sigma_{\text{act}}=10^{-5}$, and use a bisection algorithm to resolve the critical stresses at which instabilities occur to within $\delta\sigma_{\text{act}}=10^{-8}$. After each increase to $\sigma_{\text{act}}$ we evolve the system until the largest unbalanced forces are less than a threshold value of $10^{-12}$.

We characterize the plasticity in the system by calculating the magnitude of non-affine displacements $D^2_{\text{min}}$ between the stable configurations before and after each avalanche. This follows Ref.~\cite{falk_dynamics_1998}, finding the best-fit affine transformation for a neighborhood of radius $r=5$ around each particle and subtracting this from the observed displacements. Note that this procedure gives a value of $D^2_{\text{min}}$ for \emph{each individual} spherical particle making up the rods. 

Fig \ref{FigSM1}C shows the distribution of the maximum value of $D^2_{\text{min}}$ between stable configurations, which has three main peaks. The lowest peak corresponds to purely elastic motion, as particles largely do not change neighbors (Fig \ref{FigSM1}D). We consider a given deformation to be an avalanche if $D^2_{\text{min}}>1$, as this is the point at which particles actually change contacts, \mlm{highlighted by the left-most black vertical lines in Fig~\ref{FigSM1}(C,D).} The peak at the highest values of $D^2_{\text{min}}$ ($>10^4$) corresponds with very large strains that cause the entire system to flow; the maximum value of the distribution is limited by the system dimensions and thus scales with system size. These large deformations cannot be divided into localized rearrangements and so we ignore avalanches with $D^2_{\text{min}}>5000$, \mlm{highlighted by the right-most black vertical lines in Fig~\ref{FigSM1}(C,D).}

The middle peak contains avalanches that cause only a fraction of system to rearrange, so we perform the subsequent clustering analysis on avalanches with maximum values of $D^2_{\text{min}}$ between $1$ and $5000$ (black lines in Fig \ref{FigSM1}C,D).

\subsubsection{Numerically Generating Cubic Modes}
Note that a ``mode" $\mathbf{\hat{z}}$ for our model of self-propelled rods consists of an $x$, $y$, and $\theta$ component for each rod: $\mathbf{\hat{z}}_\alpha = [\hat{z}_{\alpha x}\; \hat{z}_{\alpha y}\; \hat{z}_{\alpha \theta}]^T$. However, for some purposes (such as visualization) it is useful to convert this into purely translational components for each of the particles making up a rod:
\begin{align}
    \mathbf{\hat{z}}_{\alpha i} = \left[\begin{array}{c}
          \hat{z}_{\alpha x} - \hat{z}_{\alpha \theta}\, \text{sin}\,\theta_\alpha \\
          \hat{z}_{\alpha y} + \hat{z}_{\alpha \theta}\, \text{cos}\,\theta_\alpha
    \end{array}\right].\label{rod mode to particle mode}
\end{align}
We also make this conversion when calculating the participation ratio 
\begin{align}
    P_r(\mathbf{\hat{z}}) = \frac{\left(\sum_{\alpha i} |\mathbf{\hat{z}}_{\alpha i}|^2 \right)^2}{N \left(\sum_{\alpha i} |\mathbf{\hat{z}}_{\alpha i}|^4\right)}
\end{align}
as shown in Fig 2E in the main text.

For each stable configuration, we construct the tensors $\tilde{\mathcal{M}}$ and $\tilde{\mathcal{T}}$ using analytical expressions for the derivatives of Eqn \ref{active rods forces} and obtain ensembles of harmonic and cubic modes. Expressions for these derivatives in the active rod system are included in the open source software associated with this manuscript~\cite{active-rods_GitHub}.

We calculate the harmonic modes directly by numerically diagonalizing the augmented Hessian $\tilde{\mathcal{M}}$. As the non-conservative forces make $\tilde{\mathcal{M}}$ non-symmetric, these are no longer normal vibrational modes: in fact, they are not even guaranteed to be real. However, we observe that for the majority of configurations of our model $\tilde{\mathcal{M}}$ has all real eigenvalues/eigenvectors. Around 10\% of configurations have between 1-3 complex eigenvalues, but even these are ``almost real" in that their imaginary parts are at least 4 orders of magnitude smaller than their real parts. It would be interesting for future work to consider why this is the case. However, as these slightly-complex eigenvectors are not in the low-$\kappa$ portion of the spectrum, we will not consider them further in this work, as we only utilize the lowest harmonic mode as an indicator of the unstable direction at the onset of an avalanche.

We also use the harmonic modes as a convenient set of initial conditions for the mapping in Eqn \ref{cubic mode mapping}, which we iterate until Eq \ref{cubic mode definition nc} is satisfied to within floating point precision, indicating we have found a cubic mode. \tah{To ensure a consistent population of cubic modes between configurations, we then use all previously-obtained modes for each member of the ensemble as additional initial conditions.} As many initial conditions will converge to the same fixed point, we take only the unique cubic modes produced by this procedure. While this protocol does not necessarily produce all possible cubic modes, it preferentially finds the modes with the smallest $\kappa(\mathbf{\hat{\pi}})$ and $b(\mathbf{\hat{\pi}})$, which are exactly the modes that have the greatest impact on the yielding behavior. \tah{In non-active amorphous systems, it has been shown that low-frequency soft excitations produce a ``halo effect": there is a diverging length scale $\xi\propto \omega^{-1}$ that separates low-frequency modes from other minima of the energy barrier cost function, so that very close to an instability only a single minimum exists \cite{richard_detecting_2023}.}

\begin{figure}[h!]
\centering
\includegraphics[scale= 0.121428]{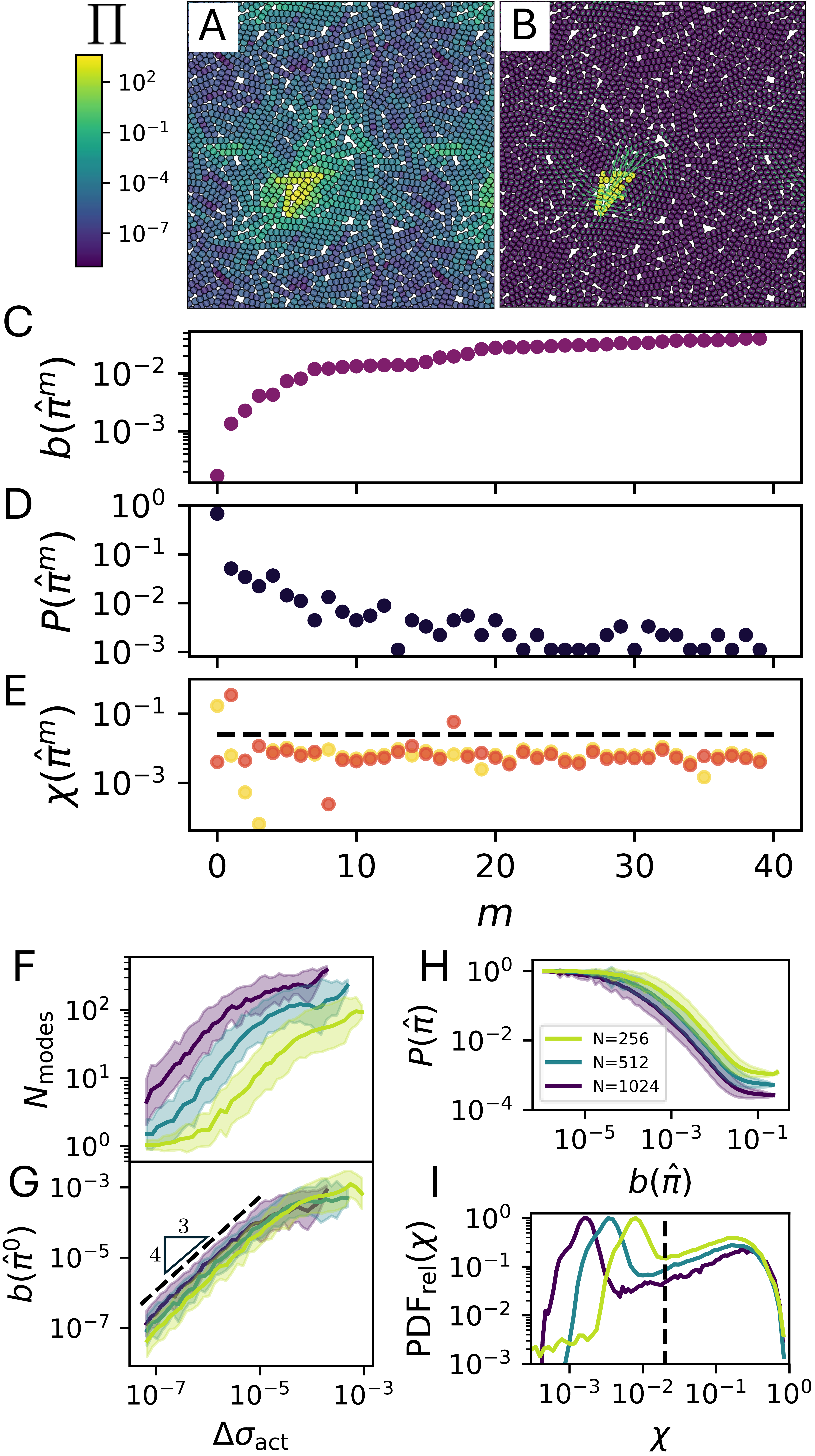}
\caption{A) Values of the scalar field $\Pi$. B) Cubic mode $\hat{\pi}$ corresponding to $\Pi$ in A), along with the associated cluster of points (same as Fig 3B in the main text). C-E) Energy barriers (C), convergence probabilities (D) and proficiencies (E) of the lowest 40 cubic modes (indexed by $m$) of the configuration shown in A-B. The proficiencies are calculated between the cubic modes and the two rearrangements shown in Fig 3D of the main text, with different colors corresponding to each rearrangement. The lowest two cubic modes each have a high proficiency (greater than the threshold $\chi_{\text{th}}=0.02$, shown with a dashed line) with one of the rearrangement clusters. \tah{(F) Average number of cubic modes obtained at different intervals of stress $\Delta\sigma_{\text{act}}$ before an instability for different system sizes (given by legend in (H)). (G) Average value of the smallest energy barrier $b(\hat{\pi}^0)$ at different $\Delta\sigma_{\text{act}}$. Dashed line indicates the scaling $b(\hat{\pi}^0)\propto (\Delta\sigma_{\text{act}})^{4/3}$. (H) Average probability of converging to a cubic mode as a function of the energy barrier $b(\hat{\pi})$. (I) Relative probability density of the proficiency $\chi$ between cubic modes and rearrangements. Dashed line indicates the threshold value $\chi_{\text{th}}$ used to separate ``successes" from ``failures".}}
\label{FigSM2}
\end{figure}

In Fig \ref{FigSM2} we show how both the number of cubic modes (F) and the smallest energy barrier $b(\mathbf{\hat{\pi}^0})$ (G) depend on the distance in stress $\Delta\sigma_{\text{act}}$ from the next instability. The closer to an impending avalanche, the lower the barrier of the lowest mode, as expected from Fig 3G in the main text. We observe the scaling relationship $b(\hat{\pi}^0)\propto (\Delta\sigma_{\text{act}})^{4/3}$. The number of cubic modes found drops to as low as a single mode for our smallest system size just before an avalanche. In this case all initial conditions converge to the same cubic mode, implying that the basin of attraction around each fixed point of Eqn \ref{cubic mode mapping} increases as $b(\mathbf{\hat{\pi}})$ decreases, as has been observed in non-active systems \cite{richard_detecting_2023}. 

In Fig \ref{FigSM2}H, we plot the average probability of converging to a cubic mode (that is, the fraction of initial conditions that converge to that mode) as a function of $b(\mathbf{\hat{\pi}})$, which supports this conclusion. In Fig \ref{FigSM2}D we show the probability of converging to each of the lowest-$b$ cubic modes of the configuration in Fig \ref{FigSM2}A-B, where it is clear that it is more probable to converge to cubic modes with lower energy barriers even in a \emph{single} configuration.

Once we have obtained a population of cubic modes, we associate each mode with a cluster of particles using persistent homology, which we describe in detail in the next subsection. We perform the clustering on a scalar field $\Pi$ for each mode $\hat{\pi}$ that describes the strength of the force response for each particle. First, the first-order force response is given by 
\begin{align}
    \left[\begin{array}{c}
        \mathbf{f}^{(1)}_{\mathbf{r}}  \\
        \mathbf{f}^{(1)}_{\theta}
    \end{array}\right]=
    \mathbf{f}^{(1)} = \tilde{\mathcal{M}}\cdot\hat{\pi} = \left[\begin{array}{cc}
        \tilde{\mathcal{M}}_{\mathbf{r}\mathbf{r}} & \tilde{\mathcal{M}}_{\mathbf{r}\theta} \\
        \tilde{\mathcal{M}}_{\theta\mathbf{r}} & \tilde{\mathcal{M}}_{\theta\theta}
    \end{array}\right]\left[\begin{array}{c}
        \hat{\pi}_{\mathbf{r}}  \\
        \hat{\pi}_{\theta}
    \end{array}\right],
\end{align}
where we have explicitly separated the augmented Hessian, cubic mode, and force response into their translational and rotational components. By definition the second-order force response $\mathbf{f}^{(2)}$ to a cubic mode is proportional to $\mathbf{f}^{(1)}$, so this quantity captures the full distribution of forces to second order up to an overall scale, which does not affect the clustering procedure. The scalar $\Pi$ is simply the magnitude of the force response on each individual particle:
\begin{align}
    \Pi_{\alpha i} = |\mathbf{f}^{(1)}_{\alpha i}|^2,
\end{align}
where (similar to Eqn \ref{rod mode to particle mode})
\begin{align}
    \mathbf{f}^{(1)}_{\alpha i} = \left[\begin{array}{c}
          f^{(1)}_{\alpha x} - f^{(1)}_{\alpha \theta}\, \text{sin}\,\theta_\alpha \\[2pt]
          f^{(1)}_{\alpha y} + f^{(1)}_{\alpha \theta}\, \text{cos}\,\theta_\alpha
    \end{array}\right].
\end{align}
In Fig \ref{FigSM2}A we visualize the field $\Pi$ for the cubic mode in Fig 3B in the main text (also included here in Fig \ref{FigSM2}B).

\subsubsection{Persistent Homology Clustering}

Following reference~\cite{stanifer_avalanche_2022}, we use persistent homology to cluster the $D^2_{\text{min}}$ fields into individual rearrangements and the cubic mode fields into discrete soft spots. Again, this procedure is done at the levels of the individual particle that comprise the rods. In this algorithm particles are sorted by the values of the scalar field and assigned to clusters in order from largest to smallest. If a given particle does not neighbor an existing cluster, it is placed in a new cluster which is said to have a "birth" value equal to the scalar field value of that particle. If a given particle neighbors a single existing cluster, it is simply added to that cluster. If a given particle neighbors two or more existing clusters, those "parent" clusters are assigned a "death" value equal to the scalar field value of the given particle and a new "child" cluster is created by combining the parent clusters and is assigned a birth value equal to the scalar field value of the given particle. 

This procedure continues until all particles have been assigned to clusters, and we are left with a tree of parent-child relationships between clusters of various sizes. We then need to prune this tree to pick out the most important clusters. First we choose a maximum and minimum size for our clusters, which we choose to be $N_{\text{min}}=5$ (the number of particles in a single rod) and $N_{\text{max}} = 100$ (approximately the number of particles in a neighborhood for calculating the $\Dmin$). All clusters with sizes not between these values are pruned from the tree.

We then want to choose the clusters that are most "persistent": the persistence of a cluster is defined as the difference between its birth value and death value. For each cluster left in the tree, we look at all of the clusters that are above it: i.e. the parents, the parents' parents, etc. If there are two or more upstream clusters that are more persistent than the child cluster, this implies that multiple significant events are being combined, so the child cluster is pruned. If fewer than two upstream clusters are more persistent than the child cluster, then the dominant feature is being captured successfully. All upstream clusters are then pruned, as they would split a single event into multiple less significant clusters.

If one does not know the structure of the data the persistence distribution can help separate signal from noise, but our case is more simple. We know that each cubic mode is quasi-localized, so we take only the single cluster containing the maximum value of $\Pi$ for each mode as the soft spot. In addition, the distribution of the maximum $\Dmin$ value between stable configurations (Fig \ref{FigSM1}C) shows a separation between purely elastic deformation ($\Dmin<1$) and avalanches ($\Dmin>1$). As we observe the smallest avalanches involve only one plastic event, we take $\Dmin=1$ to also be the lower bound on the $\Dmin$ in a single rearrangement, and so we only select the clusters produced by persistent homology that contain a particle with $\Dmin>1$. 

We can also use this bound to order the clusters within an avalanche. While the persistent homology clustering is performed on the final $\Dmin$ field calculated between the initial and final stable configurations, we also calculate the $\Dmin$ at multiple additional points during the relaxation. We then take the time at which the maximum value of $\Dmin$ in a given cluster passes the threshold $\Dmin=1$ as the ``onset time" of that cluster.

\subsubsection{Proficiency}

To quantify the correlation between a cubic mode and a future rearrangement, following Ref.~\cite{stanifer_avalanche_2022}, we calculate the proficiency $\chi$ between their corresponding clusters of points. We represent a cluster of points by a binary field $c_{\alpha i}$, which is 1 if particle $i$ of rod $\alpha$ is in the cluster and 0 otherwise. Then the mutual information between two clusters $c^1$ and $c^2$ is \cite{stanifer_avalanche_2022}
\begin{align}
    I(c^1, c^2) = \sum_{x,y\in[0,1]} P_{(c^1,c^2)}(x,y) \,\text{log}\left(\frac{P_{(c^1,c^2)}(x,y)}{P_{c^1}(x)P_{c^2}(y)}\right),
\end{align}
where $P_{c^1}(x)$ is the marginal probability that $c^1=x$ and $P_{(c^1,c^2)}(x,y)$ is the joint probability that $c^1=x$ and $c^2=y$. For a given pair of clusters these are found by counting the fraction of particles for each condition: for example, $P_{c^1}(1)$ is the fraction of particles in cluster 1, and $P_{(c^1,c^2)}(0,1)$ is the fraction of particles that are both in cluster 2 and not in cluster 1. As this quantity measures overlaps between the clusters as well as their complements, it is more robust than simply counting how many particles are shared between clusters.

The proficiency between a cubic mode cluster $c^m$ and a rearrangement cluster $c^r$ is simply the mutual information normalized by the entropy of the rearrangement cluster
\begin{align}
    \chi(c^m, c^r) = \frac{I(c^m, c^r)}{H(c^r)},
\end{align}
where the entropy is given by
\begin{align}
    H(c^r) = -\sum_{x\in[0,1]} P_{c^r}(x) \,\text{log}\left(P_{c^r}(x)\right).
\end{align}
$\chi(c^m, c^r)$ represents the fraction of information in the rearrangement cluster that can be explained by the cubic mode cluster.

\tah{In Fig \ref{FigSM2}I we show the distribution of proficiencies measured between all cubic modes and all rearrangements for each system size. The distribution is bimodal, so we choose a threshold value $\chi_{\text{th}} = 0.02$ to separate ``successful" predictions from failures, which is similar to thresholds used in prior work \cite{stanifer_avalanche_2022}.}

\tah{As discussed in the main text, our procedure can produce hundreds of cubic modes for a single configuration, which makes it likely that at least one mode will have a high proficiency with any rearrangement purely by chance. Instead, we limit ourselves to only the first $N_m$ cubic modes, ordered by the energy barrier $b(\hat{\pi})$. To choose $N_m$ we first measure the baseline probability $C_0(N_m)$ that at least one out of $N_m$ Poisson-distributed random clusters (with the same size statistics as the clusters derived from cubic modes) has a high proficiency (greater than the threshold value $\chi_{\text{th}} = 0.02$) with a rearrangement. We plot this baseline rate in Fig3E, which collapses onto a single curve when rescaled by system size. We note that using different values for the threshold $\chi_{\text{th}}$ results in a worse collapse here, lending some additional credence to our choice.}

\tah{As discussed in the main text, we measure for each of our ensembles the probability $C(N_m, N_r)$ that one of the $N_m$ lowest-$b$ cubic modes of a configuration has a high proficiency (greater than a threshold value $\chi_\text{th}=0.02$) with the $N_r$'th upcoming rearrangement. In Fig 3A we plot $C$ for several values of $N_m$, and in Fig 3B we show the increase in successful predictions relative to the baseline $C-C_0$. By averaging this over all values of $N_r$ we obtain a metric for how much better than random chance the cubic modes can predict rearrangements when using different numbers of modes, which we plot in Fig 3C. We see that for all system sizes there is a clear optimal choice of $N_m$. Interestingly, while one might expect this to scale with the system volume $\propto N$, we observe that it instead scales with the linear system dimension $\propto \sqrt{N}$. Based on this, we choose $N_m = 10, 15, 20$ for $N=256,512,1024$ respectively, which are indicated by dashed lines in Fig 3D-E. \tah{Recall that when close to an instability, the number of cubic modes located by our protocol decreases (Fig \ref{FigSM2}F). For our analysis of $C$, we only consider configurations where we find at least $N_m$ cubic modes.} These numbers of modes cover no more than $\approx 20\%$ of the system, which can be seen in Fig 3D where we plot the average fraction of particles $\Phi$ that are contained in $N_m$ clusters. \mlm{This is comparable to the coverage of defects found in other particle-based systems which are ductile (i.e. have not been specially prepared to be ultrastable)~\cite{manning_vibrational_2011, richard_predicting_2020}.}}

\begin{figure}[ht!]
\centering
\includegraphics[scale= 1]{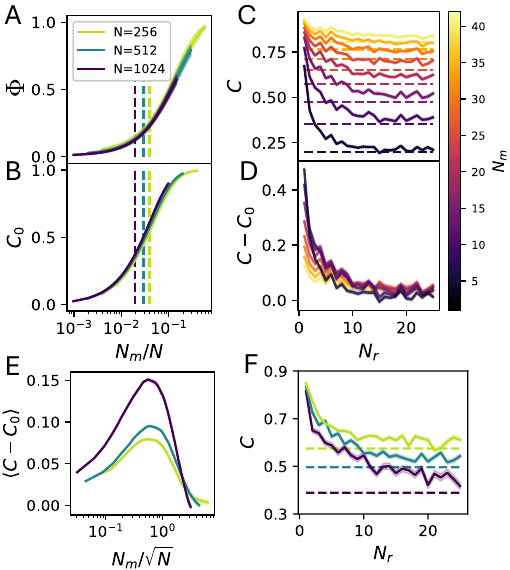}
\caption{\tah{(A) Average fraction of particles $\Phi$ that are contained in $N_m$ clusters for different system sizes (legend also applies to B,E,F). (B) Baseline probability $C_0$ for different $N_m$ and system sizes. (C) Probability $C$ of possessing a high-proficiency cubic mode with different choices of $N_m$ shown in different colors. Dashed lines indicate $C_0$ for each $N_m$. Data is from $N=512$. (D) Increase in successful predictions over baseline for data in C. (E) Average increase in successful predictions over baseline for different $N_m$. Note the scaling with $\sqrt{N}$. (F) The same data as Fig3I in the main text but without the rescaling of $C$. Dashed lines indicate $C_0$ for the chosen value of $N_m$ at each system size.}}
\label{FigSM3}
\end{figure}

% \bibliographystyle{unsrt}
% \bibliography{ActiveRodsBib}

\end{document}